\documentclass[aip,graphicx, amsmath,amssymb,reprint]{revtex4-1}
\usepackage{graphicx}
\draft 
\usepackage{mhchem}
\usepackage{filecontents}
\usepackage{subcaption}
\usepackage{comment}
\usepackage{svg}
\usepackage{cleveref}
\usepackage{inputenc}
\usepackage{amsmath}
\usepackage[T1,T2A]{fontenc}
\usepackage{booktabs} 
\DeclareUnicodeCharacter{00A0}{ }

\begin{document}


\title{Picosecond all-optical switching of Co/Gd based synthetic ferrimagnets } 



\author{Pingzhi Li$^*$}
\email[]{p.li1@tue.nl}
\affiliation{Department of Applied Physics, Eindhoven University of Technology, P. O. Box 513, 5600 MB Eindhoven, The Netherlands}

\author{Thomas J. Kools}
\affiliation{Department of Applied Physics, Eindhoven University of Technology, P. O. Box 513, 5600 MB Eindhoven, The Netherlands}

\author{Hamed Pezeshki}
\affiliation{Department of Applied Physics, Eindhoven University of Technology, P. O. Box 513, 5600 MB Eindhoven, The Netherlands}

\author{Joao M. B. E. Joosten}
\affiliation{Department of Applied Physics, Eindhoven University of Technology, P. O. Box 513, 5600 MB Eindhoven, The Netherlands}

\author{Jianing Li}
\affiliation{Department of Applied Physics, Eindhoven University of Technology, P. O. Box 513, 5600 MB Eindhoven, The Netherlands}

\author{Junta Igarashi}
\affiliation{Institut Jean Lamour, UMR CNRS
7198 – Université de Lorraine – Boulevard des Aiguillettes, BP 70239, Vandoeuvre cedex F-54506, France}

\author{Julius Hohlfeld}
\affiliation{Institut Jean Lamour, UMR CNRS
7198 – Université de Lorraine – Boulevard des Aiguillettes, BP 70239, Vandoeuvre cedex F-54506, France}

\author{Reinoud Lavrijsen}
\affiliation{Department of Applied Physics, Eindhoven University of Technology, P. O. Box 513, 5600 MB Eindhoven, The Netherlands}

\author{Stephane Mangin}
\affiliation{Institut Jean Lamour, UMR CNRS
7198 – Université de Lorraine – Boulevard des Aiguillettes, BP 70239, Vandoeuvre cedex F-54506, France}

\author{Gregory Malinowski}
\affiliation{Institut Jean Lamour, UMR CNRS
7198 – Université de Lorraine – Boulevard des Aiguillettes, BP 70239, Vandoeuvre cedex F-54506, France}

\author{Bert Koopmans}
\affiliation{Department of Applied Physics, Eindhoven University of Technology, P. O. Box 513, 5600 MB Eindhoven, The Netherlands}


\date{\today}

\begin{abstract}

Single pulse all-optical switching of magnetization (AOS) in Co/Gd based synthetic ferrimagnets carries promises for hybrid spintronic-photonic integration. 
A crucial next step progressing towards this vision is to gain insight into AOS and multi-domain state (MDS) behavior using longer pulses, which is compatible with state-of-the-art integrated photonics. In this work, we present our studies on the AOS and MDS of [Co/Gd]$_{n}$ ($n$ = 1, 2) using ps optical pulses across a large composition range. 
We theoretically and experimentally show that a large Gd layer thickness can enhance the AOS energy efficiency and maximum pulse duration. We have identified two augmenting roles of Gd in extending the maximum pulse duration. On the inter-atomic level, we found that more Gd offers a prolonged angular momentum supply to Co. On the micromagnetic level, a higher Gd content brings the system to be closer to magnetic compensation in the equilibrized hot state, thereby reducing the driving force for thermally assisted nucleation of domain walls, combating the formation of a MDS. Our study presents a composition overview of AOS in [Co/Gd]$_n$ and offers useful physical insights regarding AOS fundamentals as well as the projected photonic integration.

\end{abstract}


\maketitle 


\section{Introduction}

Single-pulse all-optical switching of magnetization (AOS) in Gd-based metallic ferrimagnets\cite{Stanciu:2007aa,Ostler:2012aa,Radu:2011aa}, has attracted significant interests within the community for its application potential in memory and storage, owing to its ultrafast switching speed and very low energy dissipation\cite{Kimel:2019aa, Kim:2022aa}. Initially discovered in GdFeCo alloys\cite{Ostler:2012aa}, AOS was later also found in multi-layer synthetic ferrimagnets, such as Co/Gd\cite{Lalieu:2017aa} and [Co/Tb]$_n$\cite{Aviles-Felix:2019aa}. A unique merit of Co/Gd based synthetic ferrimagnets is their large composition (layer thickness) range for low threshold energy AOS\cite{Lalieu:2017aa, Beens:2019aa, Li:2022ac,Li:2021wr} in comparison with CoGd based alloys\cite{Xu:2017aa,Beens:2019aa,Wei:2021ui} and [Co/Tb]$_n$\cite{Peng:2023aa,Aviles-Felix:2019aa}. This difference has been attributed to a different switching mechanism\cite{Beens:2019aa,Gerlach:2017aa}, related to the notion that in [Co/Gd]$_n$, AOS is suggested to be driven by interfacial exchange scattering. Moreover, the switching speed of AOS in Co/Gd\cite{Peeters:2022uo} was found to be similar to that of CoGd alloys\cite{Radu:2011aa, Xu:2017aa} , which displays a zero crossing of the magnetization within 1 ps.
[Co/Gd]$_n$ is also wafer-scale production compatible and allows for additional interface engineering, such as combining with a possibility of adding Tb (Gd/Co/Tb) for improved anisotropy without losing the fast switching speed of AOS\cite{Hintermayr:2023aa}. 

 In recent years, interests involving integrating AOS in spintronics has naturally emerged\cite{Kimel_AOS_Review2019, Sobolewska:2020aa, Becker:2017aa}. One approach is to create AOS compatible magnetic tunnel junctions (AOS-MTJ), in which the free-layer can be all-optically switched. Synthetic ferrimagnets-based AOS-MTJs, such as those using Co/Gd\cite{Luding:2022ur} and [Co/Tb]$_n$\cite{Salomoni:2023aa} as free layer, have yielded significantly better tunneling magneto-resistance than the GdFeCo alloy counterpart\cite{Chen:2017aa}. Meanwhile, another typical attempt is to interface integrated photonics monolithically with spintronic racetrack memory\cite{Lalieu:2019aa,Li:2022ac,Pezeshki:2022aa,Pezeshki:2023aa} for direct transport of information from the photonic domain to the magnetic domain. Recently, progress demonstrating AOS in the integrated photonics circuits has been reported\cite{PingzhiThesis:2024}.

As for compatibility of AOS with integrated photonic platforms, fs pulses (as typically used in research) impose significant challenges. In particular, the large transient electric field induces strong non-linear absorption\cite{Pezeshki:2023aa,Sobolewska:2020aa,Bente:2021th,pezeshki2024integrated}. Moreover, the short pulse duration is accompanied by large dispersion, jitter and modal competitions, which impede the use of fs pulses for data transport in integrated photonics. In contrast, ps pulses can be handled by passive waveguides with high pulse energy as well as generated with high energy efficiency using on-chip laser sources\cite{Van-Gasse:2019aa}. 

\begin{figure}[t]
\centering
\hspace*{-1.0cm} 
\includegraphics[width=1.0\linewidth]{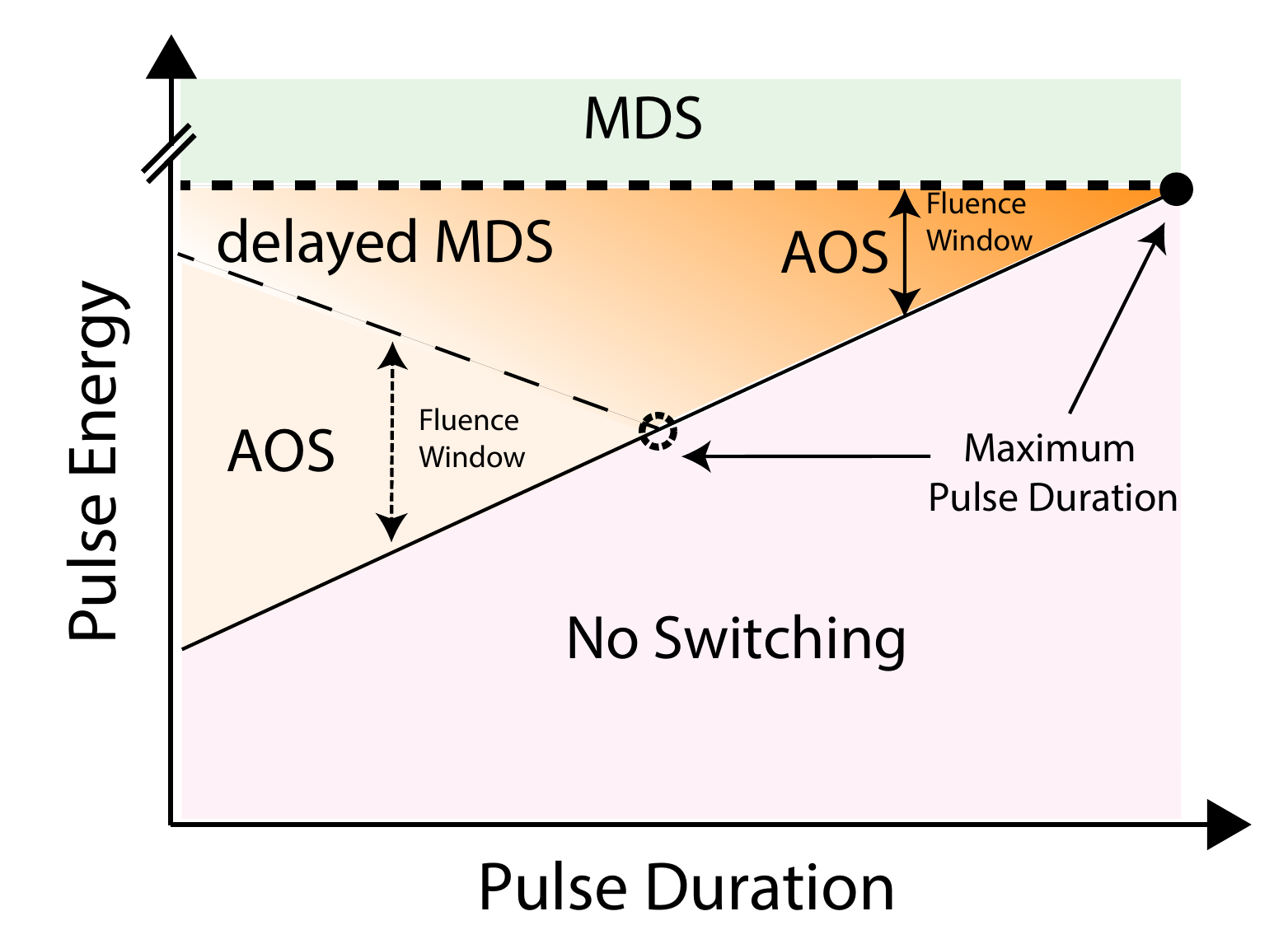}
\caption{A conceptual illustration of the phase diagram of all-optical switching (AOS), thermal demagnetization or multi-domain states (MDS) and delayed MDS as a function of pulse duration and energy. The trend of the threshold fluence of AOS and MDS (and delayed-MDS) is indicated by a solid line and a dashed line respectively. The intersection between the solid and dashed lines indicate the maximum pulse duration for AOS. }
\label{fig:ToyPhaseDiagram}
\end{figure}

Another important consideration regards the undesired occurrence of a multi-domain state (MDS) that typically occurs at fluences well above the AOS threshold\cite{Stanciu:2007aa,Ostler:2012aa,Radu:2011aa,Lalieu:2017aa}. To establish deterministic performance of devices, the occurrence of MDS formation should be avoided. In practice, this provides an incentive to optimize the fluence window, defined as the difference between threshold fluences of MDS and that of AOS at a certain pulse duration (see a conceptual illustration in FIG. \ref{fig:ToyPhaseDiagram}). In work on alloys, it has been reported that the fluence window strongly decreases upon increasing the pulse duration\cite{Wei:2021ui,Zhang:2022aa,Verges:2024aa} (see FIG. \ref{fig:ToyPhaseDiagram}). This behavior can be explained easily by the notion that AOS intrinsically requires a strong non-equilibrium between electron and phonon temperatures to occur\cite{Beens:2019aa, Mentink:2012aa,Jakobs:2021aa}. Using longer pulses, this non-equilibrium reduces, therefore leading to a higher AOS threshold energy. In contrast, in the case of MDS, the threshold energy is theoretically\cite{Jakobs:2021aa,Wei:2021ui,Zhang:2022aa} and experimentally\cite{Verges:2024aa,Wei:2021ui,Zhang:2022aa} shown to be bound by the pulse energy at which the phonon temperature transiently exceeds the Curie temperature, a process called thermal demagnetization. This MDS threshold remains almost flat for a low heat conductive substrate. Consequently, a right triangular region of AOS is resulting in a phase diagram, 
in which the fluence window linearly decreases as a function of pulse duration. A narrow fluence window can be problematic for device operation, particularly in cases of nano objects in which light absorption can be strongly inhomogeneous\cite{Pezeshki:2022aa,Pezeshki:2023aa}. Thus, when aiming at ps pulse operation, a combination of large fluence window with homogenous light absorption in a spintronic nano-object lying within the fluence window is desired during the integrated photonic design. Few studies have been carried out so far on above mentioned synthetic Co/Gd using ps pulses. 


In this paper, we present our theoretical and experimental studies on AOS and MDS in synthetic ferrimagnets of [Co/Gd]$_n$ as a function of composition and pulse duration. In order to build up a physical understanding of the effect of pulse duration on the microscopic level, we conducted theoretical studies based on the microscopic-three-temperature model (M3TM)\cite{Koopmans:2010aa, Schellekens:2014aa, Beens:2019aa}. We simulated the AOS and the thermal demagnetization using longer pulses in Co/Gd multilayers in a layer-resolved fashion and constructed its phase diagram, including a special region in which MDS is occurring in a delayed fashion without entire thermal demagnetization of the material (similar to FIG. \ref{fig:ToyPhaseDiagram}). Guided by these simulations, we further conducted experiments using ps laser pulses with varying pulse duration. 
After confirming that the AOS can be achieved with ps pulses, we investigated the composition dependence. We found that 
a higher relative Gd content can both enhance the ps-AOS energy efficiency and prolong the maximum pulse duration for AOS, which is in a qualitative agreement with our theoretical studies. Furthermore, we found that a high Gd content in our [Co/Gd]$_{n}$ suppresses the energy efficiency for AOS using short pulses but significantly enhances the energy efficiency at longer pulses. In contrast, the maximum pulse duration for a Co dominated samples is significantly lower as a result of a delayed behavior of MDS compared to that of Gd dominated samples. The phase diagram of Co dominated samples deviates from a right triangular shape, which indicates extra mechanisms beyond microscopic level also contribute to their MDS formation.
Our studies reveal the composition dependence of the AOS in [Co/Gd]$_{n}$ using ps pulses, which is an essential piece of information to establish compatibility of [Co/Gd]$_n$ with integrated photonics.

\section{Theoretical Investigation}

\begin{figure*}[t]
\hspace*{-1.5cm} 
\includegraphics[width=1.10\linewidth]{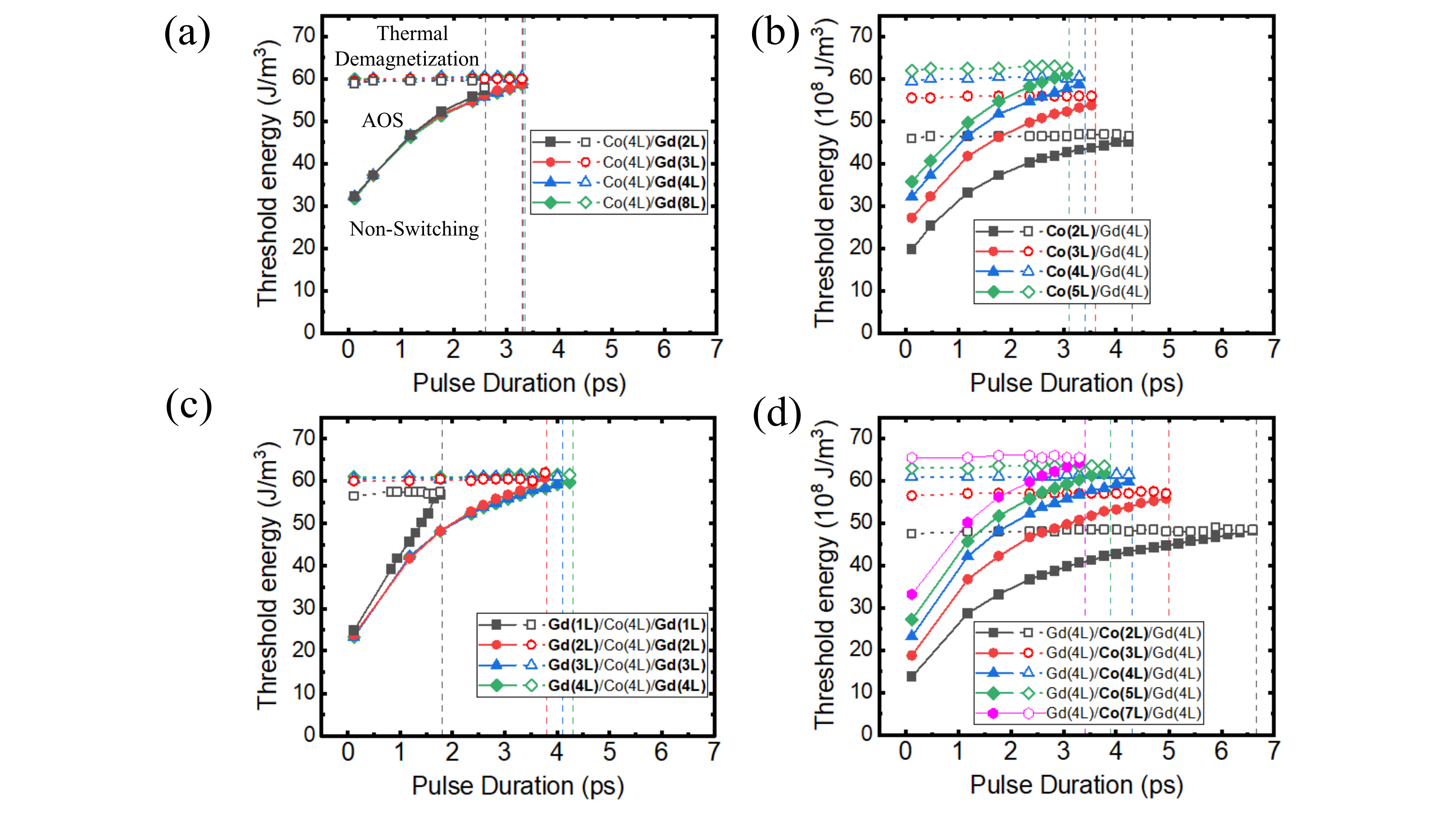}
\caption{The (absorbed) threshold energy density of AOS and thermal demagnetization (MDS) as a function of pulse duration for various composition (constituent layer thickness)  (a) Co(4L)/Gd(2, 3, 4, 8 L), (b) Co(2, 3, 4, 5L)/Gd(4L), (c) Gd(1, 2, 3, 4L)/Co(4L)/Gd(1, 2, 3, 4L), (d) Gd(4L)/Co(2, 3, 4, 5L)/Gd(4L), as obtained from our layer-resolved M3TM-based simulation. The solid symbols indicate the threshold energy for AOS. The open symbols indicate the threshold fluence for thermal demagnetization. The maximum pulse durations (within our computation resolution) for each composition were obtained from the intersection between the extrapolated trends of the two curves. Their values are marked by vertical dashed lines.}
\label{fig:M3TMPicoAOSSim}
\end{figure*}

In order to gain microscopic insight on the influence of composition on AOS and MDS, we first present our theoretical investigations on AOS and thermal demagnetization in Co/Gd multilayers. We base our theoretical approach on the M3TM with an addition of angular momentum exchange scattering between Co and Gd sublattices\cite{Koopmans:2000aa,Schellekens:2014aa}. In our model, the magnetization of each atomic layer is assumed to be a macrospin, which is governed by electron and phonon temperature dynamics. The magnetization of each layer is exchange-coupled and obtained following Weiss's mean field theory\cite{Koopmans:2010aa} (see Sup. \ref{CurieComp}). Within our model, during the transient non-adiabatic heating process induced by a short laser pulse, the angular momentum can locally be lost following the Elliot-Yafet spin-flip scattering and can be also changed via exchange scattering with the neighboring layers . We assume that the electron and phonon heat conductivity are high enough to have the electron and phonon temperature being uniform throughout our metallic stack. We treat the heat sinking to the substrate phenomenologically with a characteristic timescale of 20 ps (see Sup. \ref{M3TM}).  We keep the parameters in our simulation (see the table in Sup. \ref{M3TM}) similar to our previously presented works \cite{Koopmans:2010aa,Schellekens:2014aa,Beens:2019aa,Li:2022ab} but have adjusted the exchange scattering and Elliot-Yafet spin-flip scattering strength to match the simulated results with experimental ones within an order of magnitude. In this way, we simulate the AOS of our designated system in the most idealistic case with a computationally viable approach.

In our investigations, we simulated the switching dynamics of the Co/Gd multilayer structure at different pulse energy. Scanning the pulse energy upwards, at low energies, the simulated time traces only show a fast demagnetization followed by a remagnetization in the original direction. Above a certain threshold energy, a zero crossing of the magnetization in all atomic layers is found, followed by a remagnetization in the opposite direction (see FIG. \ref{fig:M3TMDynamics} in Sup. \ref{M3TM}), thus AOS takes place. We term such a threshold value as \textit{the threshold energy for AOS}.

At even higher pulse energies, the magnetization stays close to zero for very long time (see FIG. \ref{fig:M3TMDynamics} in Sup. \ref{M3TMphase1}). This observation is related to the fact that the phonon temperature remains above or close enough to the Curie temperature for a significant amount of time (see FIG. \ref{fig:M3TMPhase1} Sup. \ref{M3TMphase1}). In reality, it is well-known that a too high pulse energy above a certain threshold value brings the system into a MDS\cite{Lalieu:2017aa, Gorchon:2016aa, Wei:2021ui, Zhang:2022aa}. We thus associate our observation of this full thermal demagnetization to MDS. In our simulation, we define \textit{the threshold energy for the thermal demagnetization (or MDS)}, as the pulse energy above which, the averaged normalized magnetization of all Co layer remains smaller (less negative) than $-$10$\%$ in the reversed direction at 6 ps after arrival of the peak of pump pulse. This choice was motivated by the fact that, as a real system remains demagnetized for longer than 5 ps before switching, the thermal fluctuations would already eliminate a possible switching scenario at a later time or already disrupt a slightly switched process.

We start by discussing the generic behavior of the effect of the pulse duration on the threshold energy. FIG. \ref{fig:M3TMPicoAOSSim}a and \ref{fig:M3TMPicoAOSSim}b show the extracted threshold energies for AOS and MDS, i.e. a phase diagram, for a bilayer Co/Gd sample. Using that of Co(4L)/Gd(2L) as a typical example (number in parantheses indicates the number of atomic layers), the threshold energy for AOS increases as a function of the pulse duration. We attribute this behavior to the fact that prolonged pulses reduce the non-adiabatic temperature difference between electrons and phonons, thus, the peak electron temperature reduces, which is commonly considered a key criterion for AOS\cite{Mentink:2012aa, Beens:2019aa}. We also see that the threshold energy for MDS remains almost flat, which is consistent with the earlier experimental observations in GdCo-based alloys\cite{Wei:2021ui, Zhang:2022aa} and earlier atomistic-LLG simulations\cite{Raposo:2022aa, Jakobs:2021aa,Wei:2021ui, Zhang:2022aa}. 
As the pulse duration increases, it can be seen that the switching window gradually closes till the two curves intersect at the maximum pulse duration. Such a behavior is consistently observed in all compositions we have simulated.

After having verified the generic behavior, we continue our discussion on the influence of the composition on the threshold energy and maximum pulse duration. We will first address the dependence on the number of Gd layers. FIG. \ref{fig:M3TMPicoAOSSim}a displays results from different Gd thickness, while the effect of changing the Co thickness is shown in FIG. \ref{fig:M3TMPicoAOSSim}b. As compared with the bilayer composition of Co(4L)/Gd(2L),  FIG. \ref{fig:M3TMPicoAOSSim}a shows that adding one additional or more Gd layers almost does not change the threshold energy for both AOS and thermal demagnetization. Since the Curie temperature of Gd is below room temperature, we understand this by the notion that the Curie temperature is dominated by the Co layer (see Sup. \ref{CurieComp}). It can be also observed that the maximum pulse duration increases adding one additional Gd layer. Next we consider the influence of the amount of Co layers (see  FIG. \ref{fig:M3TMPicoAOSSim}b). We can see that the threshold energy for both AOS and thermal demagnetization increases upon adding more Co layers from 2 to 7, as a consequence of the associated increase of Curie temperature with added Co layers, which enhance the exchange stabilization (see Sup. \ref{CurieComp}). The maximum pulse duration displays an opposite trend, i.e. it decreases from approximately 7 to 3.5 ps with increasing number of Co layers, which will be explained in a later discussion.

We next shift our focus to the effect of an added Gd interface (see FIG. \ref{fig:M3TMPicoAOSSim}c-d). Here, we address the differences between a Co/Gd bilayer and a Gd/Co/Gd trilayer. In FIG. \ref{fig:M3TMPicoAOSSim}c, we show the threshold energy for both AOS and thermal demagnetization as a function of Gd layers sandwiching Co (4L). We see that adding Gd can both reduce the threshold energy for AOS and extend maximum pulse duration in a similar fashion to the bilayer case, however, in a more pronounced way. Comparing Co(4L)/Gd(2L) with Gd(2L)/Co(4L)/Gd(2L) (see FIG. \ref{fig:M3TMPicoAOSSim}a and c), it can be seen that, despite Gd(2L)/Co(4L)/Gd(2L) exhibiting a significant higher threshold energy for thermal demagnetization, its AOS threshold energy is lower than Co(4L)/Gd(2L). Moreover, the maximum pulse duration of Gd(2L)/Co(4L)/Gd(2L) is 50$\%$ higher. Now comparing between Gd(1L)/Co(4L)/Gd(1L) with Gd(2L)/Co(4L)/Gd(2L), it can be seen that Gd(2L)/Co(4L)/Gd(2L) displays a lower AOS threshold energy and a higher thermal demagnetization threshold due to a higher Curie temperature as well as a significantly longer max pulse duration. Adding more Gd away from the two interfaces continues this trend, which is similar to the bilayer case of Co(4L)/Gd(2-8L) (see  FIG. \ref{fig:M3TMPicoAOSSim}a), but with a significantly larger impact  (compare between  FIG. \ref{fig:M3TMPicoAOSSim} a and c). It is also worth noting that for short pulses, the AOS threshold energies of the compositions in  FIG. \ref{fig:M3TMPicoAOSSim}c are similar, while for longer pulses, those in Gd(2-4L)/Co(4L)/Gd(2-4L) are much lower.

 Lastly, we discuss the behavior of Gd(4L)/Co(X)/Gd(4L), in which X is the number of Co layers (see  FIG. \ref{fig:M3TMPicoAOSSim}b and d). FIG. \ref{fig:M3TMPicoAOSSim}d shows the dependence of threshold energy for both AOS and thermal demagnetization as a function of number of Co layers. We see that by reducing the number of Co layers, the threshold energy for both AOS and thermal demagnetization significantly reduce, while the maximum pulse duration significantly increases. This trend is similar to the behavior of corresponding bilayer Co(X)/Gd(4L) as discussed above in FIG. \ref{fig:M3TMPicoAOSSim}b. However, with an added Co/Gd interface, the AOS threshold energy is much lower than the bilayer while with similar thermal demagnetization threshold consequently the maximum pulse duration is significantly longer for the trilayer case.

Thus, based on our theoretical investigations, we can briefly summarize at this point that an added Gd content by either increasing the amount of Gd layers, reducing the amount of Co layers or adding a Co/Gd interface can increase the energy efficiency of AOS and extend the maximum pulse duration. The changes of threshold energy for AOS induced by Co are mainly due to the modification of the Curie temperature, while that induced by Gd are mainly a result of the larger angular momentum transfer from added Gd contents. For instance, the bilayer and trilayer share similar threshold energy for thermal demagnetization, but the trilayer displays a much lower threshold energy for AOS. As for the maximum pulse duration, we understand this behavior based on the fact that an increased relative Gd content effectively increases the time of the demagnetization process and the angular momentum transfer via the exchange scattering with the Gd sub-system \cite{Jakobs:2021aa, Davies:2020ab, Kim:2022aa}. During these processes, the Gd content can continuously supply reversed angular momentum to already demagnetized Co assisting it to switch sign at a lower pulse energy. Moreover, using longer pulses leads to prolonged AOS dynamics, which is more susceptible to the thermal agitation. More reversed angular momentum from Gd also helps to stabilize switching, which facilitates the AOS at longer pulses.

\section{Experimental Setup}

\begin{figure*}
\centering
\includegraphics[width=1\linewidth]{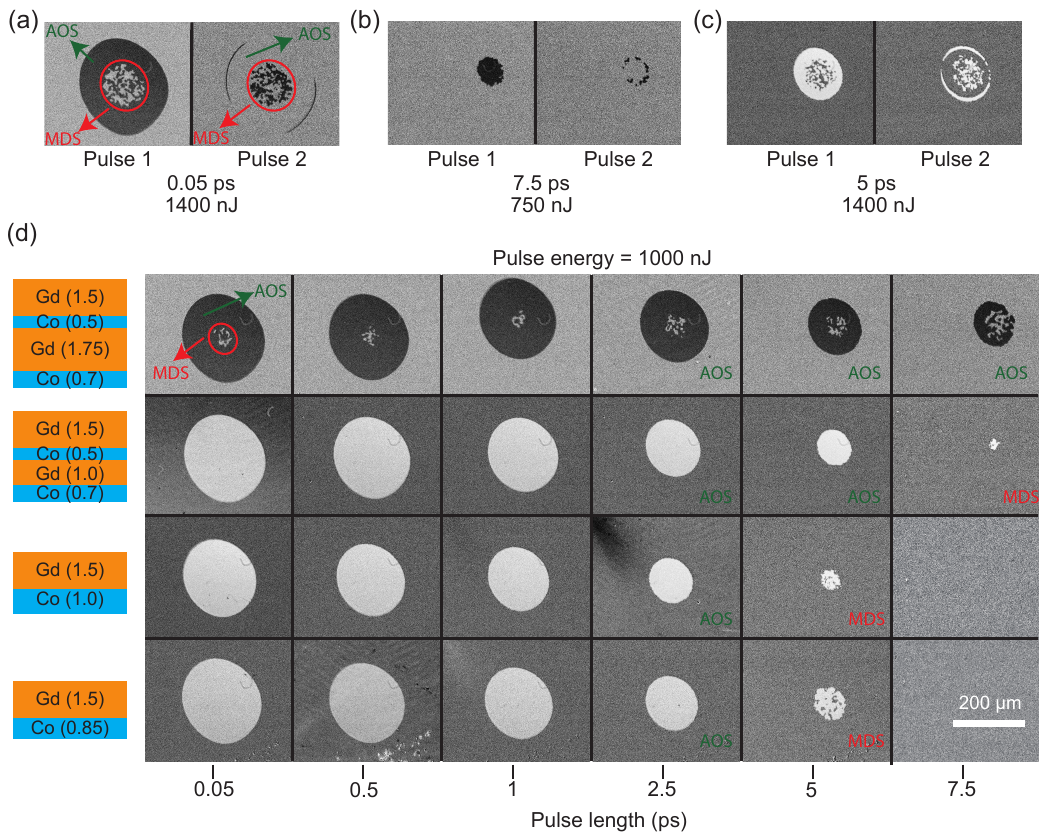}
\caption{Kerr images of switched domains of various composition, pulse energy and duration. All the images share the same scale. The identification as either AOS or MDS is of given in some of the images. A whole image is marked as AOS if not the whole area is MDS.
 (a)-(c): Kerr image after the first and second shot of the samples with various composition, pulse energy and pulse duration. (a) and (b) are for composition Co(0.7)/Gd(1.75)/Co(0.5)/Gd(1.5), (c) is for Co(0.7)/Gd(1.0)/Co(0.5)/Gd(1.5).  (d) Kerr images of the first shot for composition (shown left of the corresponding row of image) after illumination with a laser pulse of various duration at a pulse energy of 1000 nJ. }
\label{fig:Picosecond}
\end{figure*}

We now proceed with experimental investigations on AOS and MDS in [Co/Gd]$_{n}$. We used DC magnetron sputtering to create our magnetic thin film sample, in which a moving wedge shutter is incorporated during deposition to create a spatial gradient of film thickness. For this study, we have wedge deposited (thickness in nm in parentheses) Ta(4)/Pt(4)/Co(0.7)/Gd(0-2.25)/Co(0-1.5)/Gd(1.5)/TaN(5) on a Si/SiO$_2$(100) substrate at a base pressure of less than $1\times 10^{-8}$ mBar. Ta/Pt serves as a seed layer and induces a large perpendicular magnetic anisotropy (PMA)\cite{Lalieu:2017aa,Li:2022ab}, while TaN is used as a capping layer to preserve the magnetic stability of Gd\cite{Kools:2023aa, Li:2022ac, Kools:2022aa} at ambient condition. From here on, we will not mention the seed layer and capping layer when referring to the exact composition. Using a double wedge\cite{Kools:2022aa} allows us to study both the case of bilayer [Co/Gd] and quadlayer [Co/Gd]$_2$ in a single deposition run.

In our experiments, samples were illuminated by linearly polarized laser pulses with Gaussian spatial energy profile with center wavelength of 800 nm at about 45 degrees from perpendicular incidence with known pulse energy and pulse duration. As a result, AOS takes place at a certain fluence range and leaves with a switched domain. AOS was verified to take place at a ps timescale. Some examples of switching dynamics of [Co/Gd]$_{n}$ illuminated using a 0.05 ps pulse are given in FIG. \ref{fig:Fig1}g-h from Sup. \ref{CoGdBilayer}. A chirped-pulse amplifier (CPA) setup was used to stretch the pulse duration while keeping the same pulse energy. An auto-correlator was used to determine the pulse duration right after adjusting the configuration of the CPA. We exploited polar Kerr microscopy to image the switched domains at given pulse durations and energies.

Some example Kerr images of AOS and MDS following the laser illumination with the same spot size are presented in FIG. \ref{fig:Picosecond}. We determine the threshold fluence following the method commonly used\cite{Liu:1982aa,Lalieu:2017aa, Li:2021wr,Xu:2017aa,Wei:2021ui}, which is based on fitting the dependence of switched domain area on pulse energy to our Gaussian laser profile. We note that we identify a given case as switching only when we see a 100$\%$ switching after a single shot and 100$\%$ switching-back using a second shot with the same laser pulse energy at the same location as shown in FIG. \ref{fig:Picosecond}a-c. Here, we specifically note that the fringes at the edge after a second shot are due to the vibration of the sample during the measurement and pulse energy fluctuations, such that the second shot cannot fully overlap with the first shot. A MDS is identified by the appearance of more than two domains or a single domain deviating significantly from an elliptical shape in the region of illumination at a given laser fluence. Sometimes, the multi-domain region may contain more than 50\% of switched domains. In this case, we still count these cases as MDS. For example, as shown in  FIG. \ref{fig:Picosecond}d, for the MDS cases of both Co(1.0)/Gd(1.5) and Co(0.85)/Gd(1.5), with a single shot using a 5 ps pulse, the domains created are not uniform, and with a second shot, the domains cannot be switched back (not shown).

\section{Experimental Results}

\begin{figure*}
\centering
\hspace*{-1.95cm} 
\includegraphics[width=1.2\linewidth]{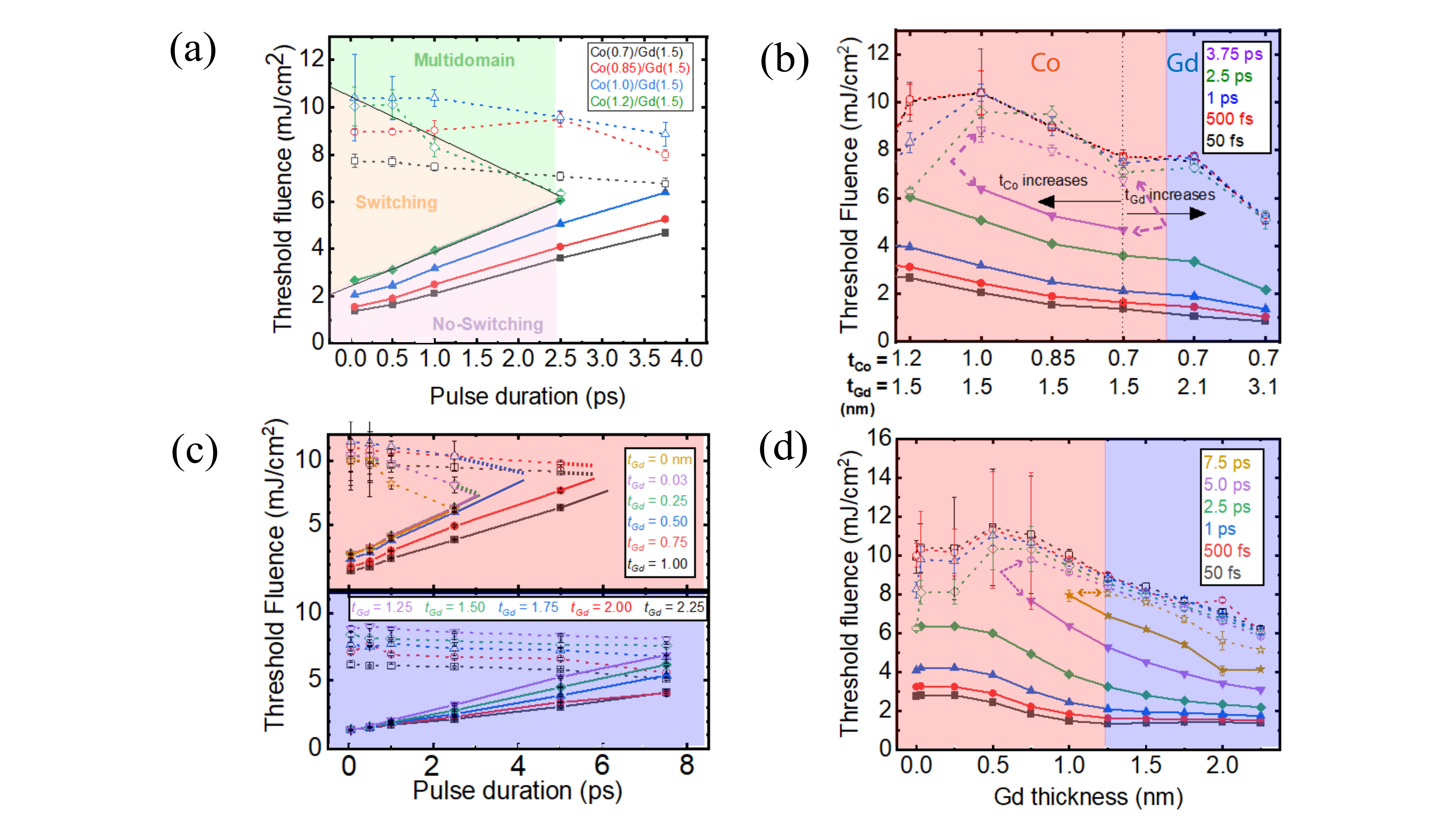}
\caption{
(a)-(d) Threshold fluence for AOS (closed symbol) and multi-domain state (open symbol) for bilayer Co(X)/Gd(Y) (a)-(b) and quadlayer Co(0.7)/Gd(X)/Co(0.5)/Gd(1.5) (c)-(d). 
(a),(c) The switching phase diagram (threshold fluence as a function of pulse duration) of bilayer Co(X)/Gd(Y) (a) and quadlayer Co(0.7)/Gd(X)/Co(0.5)/Gd(1.5) (c).
(b),(d) Composition dependence of threshold fluence for AOS and multi-domain state for bilayer Co(X)/Gd(Y) (b) and [Co/Gd]$_2$ Co(0.7)/Gd(X)/Co(0.5)/Gd(1.5) (d). 
 In (a), a color plot was made based on the phase diagram of a bilayer of Co(1.2)/Gd(1.5), indicating the region of switching (orange), non-switching (purple) and multi-domain state (green). 
 In (b), the thickness of the Co and Gd is indicated below the figure. In (c), the phase diagram of Co dominated samples are presented in the top part and Gd dominated samples are presented in the bottom part. In (b) and (d), the Co and Gd dominated regime are color coded by red and blue, respectively. Their bordering vertical line indicate the composition of magnetic compensation. A guide to the eye (in the form of colored arrow) is added in panel (b) and (d) connecting AOS and MDS threshold curves with the pulse duration. }
\label{fig:ExperimentPicoAOS}
\end{figure*}

We begin our discussion with the switching experiments on Co/Gd bilayers. We plot the threshold fluence of AOS and MDS of Co-dominated Co/Gd as a function of pulse duration in a phase diagram as shown in FIG. \ref{fig:ExperimentPicoAOS}a. A selection of results for several Co- and Gd- dominated samples are collected in FIG. \ref{fig:ExperimentPicoAOS}b, ordered from low to high Gd content, where Co- and Gd- dominated regions are accentuated by a red and blue background respectively. We first focus on the Co-dominated samples in FIG. \ref{fig:ExperimentPicoAOS}a. It can be seen that as the pulse duration increases, the AOS threshold fluence increases almost linearly, corresponding to a reduction of domain size in FIG. \ref{fig:Picosecond}d, whereas the MDS threshold almost stays constant. This makes a right triangular phase diagram. This behavior is relatively well in line with our simulation shown in FIG. \ref{fig:M3TMPicoAOSSim} as well as experiments and atomistic LLG simulation results reported in alloys\cite{Wei:2021ui, Zhang:2022aa,Verges:2024aa, Jakobs:2021aa}. The linear increase of AOS threshold fluence can be explained by the reduced non-adiabatic difference between electron and phonon temperature. In addition, in FIG. \ref{fig:ExperimentPicoAOS}b left data points, it can be observed that the threshold fluence for AOS increases with Co thickness, which is attributed to the increase of Curie temperature as our simulation predicted (see FIG. \ref{fig:M3TMPicoAOSSim}). In Sup. \ref{CoGdBilayer}, we provide more detailed composition dependence of threshold fluence  using 0.05 ps for AOS of Co/Gd bilayers.

Despite these qualitative agreements, the slight decrease of the threshold fluence for MDS as a function of pulse duration as observed in FIG. \ref{fig:ExperimentPicoAOS}a, makes the shape of the resulting phase diagram deviating from the right-triangular phase diagram. Additionally, the relative increase of MDS threshold with Co thickness is much slower than for the AOS threshold fluence, i.e. the relative fluence gap at 0.05 ps reduces from $\sim$ 6 to $\sim$ 3.5 varying Co layer thickness from 0.7 to 1.2 nm. In our simulation (see FIG. \ref{fig:M3TMPicoAOSSim}), the threshold energies for AOS and MDS almost scale proportionally, which means that the relative fluence gap stays unaltered. Moreover, we see the negative slope for MDS threshold fluence with respect to the pulse duration is largest for the largest Co layer thickness of 1.2 nm, which leads to an acute triangular phase diagram and a small value of the maximum pulse duration of ~ 2.5 ps. Our observations suggest that some other processes apart from local thermal demagnetization also play their roles in forming a MDS, as will be elaborated in the later discussion.

We have also measured Gd-dominated bilayer samples Co(0.7)/Gd(2.1, 3.1) \cite{Kools:2023aa} (see FIG. \ref{fig:ExperimentPicoAOS}b, right most region). We see that the AOS threshold fluence reduces with Gd thickness. This is in qualitative agreement with our theoretical investigation (see  FIG. \ref{fig:M3TMPicoAOSSim}a and c). Nevertheless, the maximum pulse duration for these samples is less than our Co dominated samples (except for the thickness of 1.2 nm). The reason behind this observation is not clear. As for all bilayer samples (including Co-dominated samples), we found that the maximum pulse duration for AOS is below 5 ps.

Now we shift gear to the discussion on quadlayer samples of [Co/Gd]$_2$ at composition Co(0.7)/Gd(X)/Co(0.5)/Gd(1.5). FIG. \ref{fig:ExperimentPicoAOS}c and d show the threshold fluence of both AOS and MDS of these compositions as a function of pulse duration. 
As for the dataset for 0.05 ps laser pulses shown by black symbols in FIG. \ref{fig:ExperimentPicoAOS}d, it can be seen that the threshold fluence reduces as Gd increases, approaching a minimum close to magnetization compensation (t$_{\text{Gd}}$ = 1.09 nm)\cite{Li:2022ac} (see also a zoomed-in view in Sup. \ref{CoGdBilayer} in FIG. \ref{fig:Fig1}f), and staying almost flat at the Gd dominated regime unlike the monotonic trend in bilayers as shown in FIG. \ref{fig:ExperimentPicoAOS}b. On the other hand, the multi-domain threshold fluence decreases, which we attribute to the reduction of Curie temperature as Gd dilutes the exchange coupling strength between the two Co layers. The relative fluence gap between MDS and AOS decreases from close to 8 to 5 going deeper from the compensation to Gd dominated regime. Our observation suggests that, within the Gd dominated regime in quadlayer [Co/Gd]$_2$, the added Gd content suppresses the energy efficiency of AOS using short pulses. As the pulse duration increases, interestingly, the threshold fluence increases less drastically for samples with more Gd content. Moreover, the minimum of AOS threshold fluence at compensation vanishes at longer pulses. We conclude that added Gd content is beneficial for energy efficient AOS with longer pulses.

After having explored the threshold fluence dependence in bilayer as a function of composition, we focus on the phase diagram of [Co/Gd]$_2$ Co(0.7)/Gd(X)/Co(0.5)/Gd(1.5) quad-layers. For a Gd layer thickness of 0.25 nm and below, we found that the MDS threshold fluence decreases rapidly with pulse duration, forming an “acute triangular” phase diagram (see  FIG. \ref{fig:ExperimentPicoAOS}c and FIG. \ref{fig:ToyPhaseDiagram}) similar to that of the bilayer with large Co content Co(1.2)/Gd(1.5) (see  FIG. \ref{fig:ExperimentPicoAOS}a). At higher Gd layer thickness, the phase diagram becomes a familiar “right triangle”. Interestingly, as the Gd thickness is above 0.5 nm, the maximum pulse duration for AOS in this case can be extended to more than 5 ps, which is longer than the maximum pulse duration for all our Co/Gd bilayers. Moreover, at the Gd dominated regime (see  FIG. \ref{fig:ExperimentPicoAOS}c bottom panel), the maximum pulse duration can be extended beyond 7.5 ps.    We attribute such an enhancement to the added Gd content induced by an extra interface and more Gd content away from the interface, qualitatively in line with our theoretical investigation. Likewise, various other theoretical studies\cite{Davies:2022ab, Jakobs:2021aa} also show that a higher Gd content in CoGd alloys will facilitate a longer pulse AOS. Thus, we argue that a higher Gd content is also crucial for extending the maximum pulse duration for our synthetic [Co/Gd]$_2$ structure.  It is worth mentioning that the AOS thresholds and fluence windows of Gd dominated samples at 5 ps (around $50\&$ of its AOS threshold fluence) can satisfy the requirement of AOS imposed by our integrated photonic design\cite{Pezeshki:2023aa, pezeshki2024integrated} without suffering from significant non-linear absorption losses.

\section{Discussion on MDS formation}

\begin{figure}
\centering
\hspace*{-1.2cm} 
\includegraphics[width=1.2\linewidth]{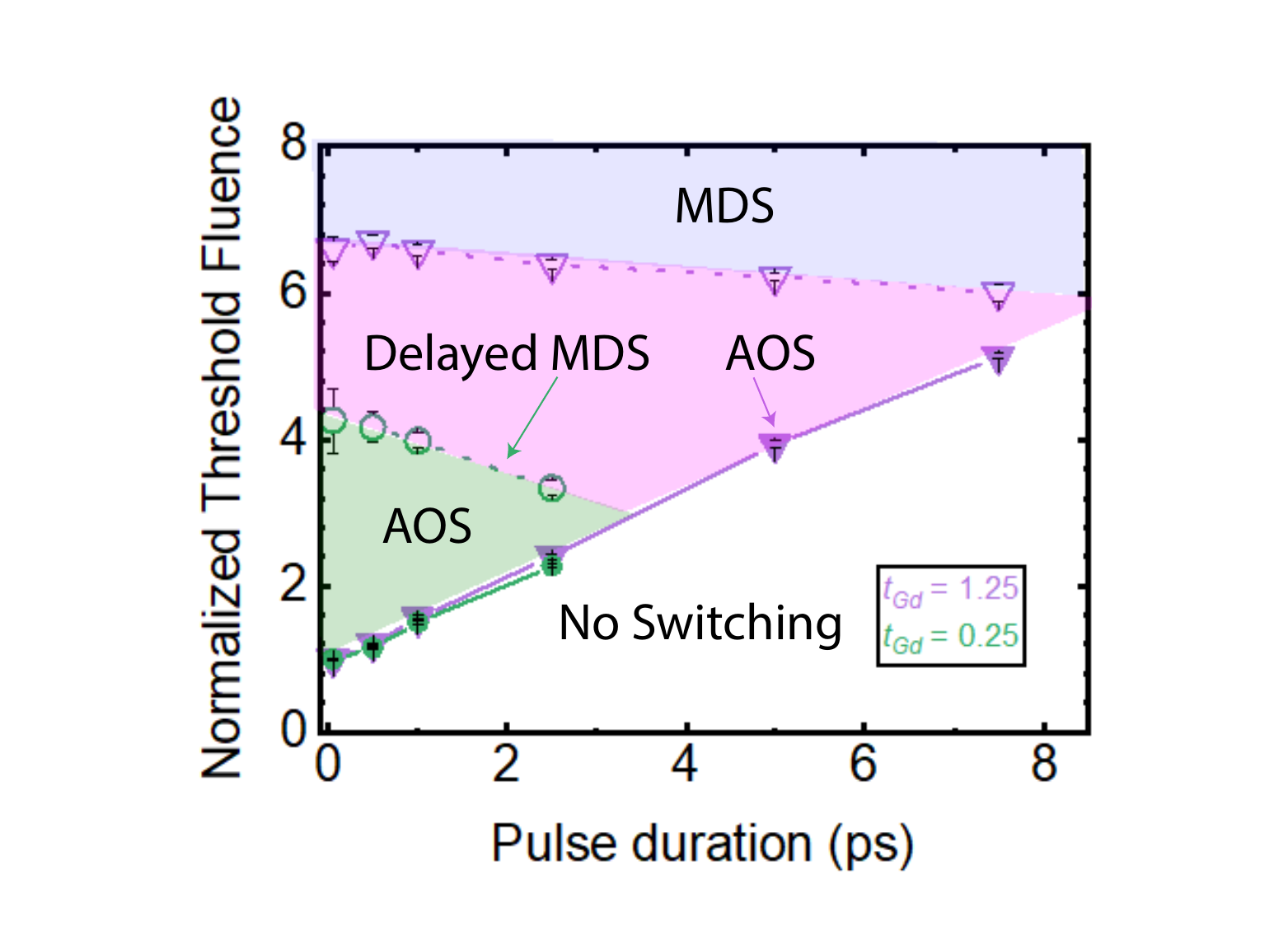}
\caption{ The phase diagram for the quadlayer sample [Co/Gd]$_2$ Co(0.7)/Gd($\textbf{0.25}$)/Co(0.5)/Gd(1.5) ($t_{\text{Gd}} = 0.25$ nm) and Co(0.7)/Gd($\textbf{1.25}$)/Co(0.5)/Gd(1.5) ($t_{\text{Gd}} = 1.25$ nm). The threshold fluence of each is normalized to their respective AOS threshold fluence using 0.05 ps pulses.
 }
\label{fig:NormalizedPhaseDiagram}
\end{figure}

It is widely accepted that the primary factor leading to the formation of a MDS is the elevation of the phonon temperature beyond the Curie temperature, a viewpoint supported by references\cite{Beens:2019aa,Jakobs:2021aa,Mentink:2012aa,Davies:2020ab, Kim:2022aa,Wei:2021ui, Gorchon:2016aa}. This aligns with our M3TM-based model as well. However, based on our experimental phase diagrams (see FIG. \ref{fig:ExperimentPicoAOS}a and c), we have observed regions within samples with high Co contents, in which MDS appears at fluences well below that required to overheat phonons above the Curie temperature. This is clearly visible by the normalized phase diagram shown in FIG. \ref{fig:NormalizedPhaseDiagram}, which shows the data from two characteristic samples with Co domination ($t_{\text{Gd}} = 0.25$ nm) and Gd domination ($t_{\text{Gd}} = 1.25$ nm) after scaling the respective fluences such that the AOS threshold fluences overlap. This rescaling accounts for the fact that the Co dominated sample has a much higher Curie-temperature. since AOS and MDS threshold fluences are expected to scale in the same way with the Curie temperature (see our simulation results in FIG. \ref{fig:M3TMPicoAOSSim}), it would have been anticipated that after rescaling also the MDS threshold for the two data sets would overlap. However, the observed trend is strongly in contrast with the prediction. More specifically, the normalized MDS threshold for the Co dominated sample is significantly smaller than expected, leading to a switching window that is reduced by a factor of almost 2 and leaving behind an extra region of MDS, as compared with the Gd dominated sample. Recent studies by Verges \textit{et al.}\cite{Verges:2024aa}, Lin\textit{et al.}\cite{Lin:2023aa} and Wei \textit{et al.}\cite{Zhang:2024aa} have suggested that the MDS formation in perpendicularly magnetized alloys traces its origin also in the influence of dipolar field.
Hereby, we conjecture that this region corresponds to a delayed formation of a MDS, well after the initial occurrence of successful AOS. This transition is driven by thermally assisted processes\cite{Lin:2023aa, Zhang:2024aa} while still being in the hot state – albeit well below the Curie temperature. 

The reason for such a delayed MDS formation at fluences where the lattice temperature remains well below the Curie temperature can be understood as follows. The completion of the AOS process results in a uniform switched domain. For this uniform state to remain stable at a longer timescale, it is required that the domain wall energy is higher than the dipolar energy of the switched region. However, as for Co dominated samples, the dipolar energy is high and effective anisotropy is low, which leads to a low domain wall energy as well as a high imbalance between the dipolar energy and domain wall energy. Right after the AOS is completed, the system still remains in a hot state, in which the relative Gd content becomes even less due to the steeper decay of Gd magnetization as a function of temperature as compared to Co. This further exacerbates this imbalance, making the system to reduce its dipolar energy by forming an MDS, nucleated by thermally-assisted processes at the elevated temperature. Moreover, with longer pulses, the degree of switching extending to the opposite side becomes less\cite{Zhang:2022aa}. This creates an initial condition for thermalization closer to MDS, which might explain the negative slope of MDS threshold fluence (see FIG. \ref{fig:NormalizedPhaseDiagram} for example). While conditions for this delayed MDS formation are optimally met for Co dominated samples, they are not for Gd dominated samples. Right after AOS, Gd dominated system may transiently become Co dominated, again due to the stronger decay of Gd magnetization with temperature. Upon cooling down, the system will remain for a long time near the compensation temperature while returning from the Co to the Gd dominated state. During this phase, the dipolar energy is much lower compared to the domain wall energy, which can much easier stabilize a uniform domain. Thus, we conclude that the delayed transition to a MDS should preferentially occur in Co-dominated samples, in line with our experimental findings. Here, it is noteworthy to mention that a full understanding of the domain formation process requires a thorough micromagnetic study including thermally assisted nucleation events, which is beyond the scope of this paper.

\section{Conclusion}

In conclusion, in this study, we theoretically and experimentally investigated the AOS and MDS formation behavior of [Co/Gd]$_{n}$ in a large composition range using ps pulses. Based on our M3TM studies, we have confirmed the augmenting role of the relative Gd content on the inter-atomic level to enhance the energy efficiency and prolong the maximum pulse duration for AOS. Our experiments on [Co/Gd]$_{2}$ show that an added Gd content can improve the energy efficiency of AOS at ps pulse duration, despite suppressing the energy efficiency of AOS at short pulses. It stabilizes a right triangular shaped phase diagram and extends the maximum pulse duration for AOS beyond 7 ps. In contrast, deep Co dominated samples exhibit an acute triangular shaped phase diagram, in which the MDS threshold energy is much lower than required for the phonon temperature to exceed the Curie temperature. We argue that the origin lies in the competition between dipolar and domain wall energy after the AOS dynamics.  This, on the other hand, suggests another augmenting role of Gd on the micromagnetic level.
Our study offers fresh insights in ps-AOS and likewise provides with guidance on designing the Co/Gd based out-of-plane magnetic stack for energy efficient all-optically switchable devices in hybrid integration between spintronics and integrated photonics.

\begin{acknowledgments}
This project has received funding from the European Union’s Horizon 2020 research and innovation programme under the Marie Skłodowska-Curie grant agreement No.860060. This work was part of the research program Foundation for Fundamental Research on Matter (FOM) and Gravitation program “Research Center for Integrated Nanophotonics,” which are financed by the Dutch Research Council (NWO). This work was supported by the Eindhoven Hendrik Casimir Institute (EHCI). This work is supported by the ANR-20-CE09-0013 UFO, ANR-20-CE24-0003 SPOTZ, and ANR-23-CE30-0047 SLAM, the Institute Carnot ICEEL, the R\'{e}gion Grand Est, the Metropole Grand Nancy for the project ``OPTIMAG'' and FASTNESS, the interdisciplinary project LUE ``MAT-PULSE'', part of the French PIA project ``Lorraine Universit\'{e} d'Excellence'' reference ANR-15-IDEX-04-LUE. This work was supported by the ANR through the France 2030 government grants EMCOM (ANR-22-PEEL-0009) and PEPR SPIN (ANR-22-EXSP-0002). J.I. acknowledges support from JSPS Overseas Research Fellowships.

\end{acknowledgments}

\section*{Author Contribution}

\textbf{Pingzhi Li}: Conceptualization (lead); Data Curation (lead); Formal Analysis (lead); Investigation (equal); Validation (lead); Methodology (equal); Visualization (lead); Supervision (equal); Writing - original draft (lead); Writing - review editing (equal);

\textbf{Thomas Kools}:  Methodology (equal); Investigation (Supporting); Writing - original draft (supporting); Writing - review editing (supporting);

\textbf{Hamed Pezeshki}:  Conceptualization(supporting); Methodology (supporting); Writing - original draft (supporting); Writing - review editing (supporting); 

\textbf{Joao Joosten}:  Methodology (supporting); Writing - original draft (supporting); Writing - review editing (supporting); 

\textbf{Jianing Li}:  Methodology (supporting); Writing - original draft (supporting); Writing - review editing (supporting); 


\textbf{Julius Hohlfeld}:  Investigation (Supporting); Writing - review editing (supporting); 

\textbf{Junta Igarashi}:  Methodology (supporting);  Investigation (Supporting); Writing - review editing (supporting); 

\textbf{Reinoud Lavrijsen}:  Funding Acquisition (equal); Writing - original draft (supporting); Writing - review editing (supporting); 

\textbf{Stephane Mangin}: Methodology(supporting); Investigation (supporting); Validation (supporting); Supervision (supporting); Funding Acquisition (supporting);  Writing - review editing (supporting);

\textbf{Gregory Malinowski}: Methodology(equal);Investigation (equal); Validation (supporting); Supervision (equal); Funding Acquisition (equal); Writing - original draft (supporting); Writing - review editing (supporting);

\textbf{Bert Koopmans}: Investigation (equal); Supervision (equal); Funding Acquisition (equal); Validation (supporting); Writing - original draft (supporting); Writing - review editing (equal);



\bibliography{LibraryThesis}

    \clearpage
    \newpage

\appendix

\section{Supplementary information}

\subsection{Sup: Magneto-static estimate of Co/Gd using Weiss's mean field model}\label{CurieComp}

In this section, we provide a simplified theoretical insight into the magneto-statics of Co/Gd. As demonstrated in Figure \ref{fig:ExperimentPicoAOS}b (and Figure \ref{fig:Fig1}b), we observe a phenomenon where the magnetism induced in Gd can potentially surpass that of Co in specific compositions of the Co/Gd system. We aim to theoretically elucidate how the magnetism in Gd, initially induced from a pristine Co/Gd interface, evolves in a static context with varying layer numbers. To quantify this evolution, we employ exchange splitting defined by applying Boltzmann statistics, based on an fcc oriented magnetic structure, following:

\begin{equation}\label{eq:exchangesplitting}
    \Delta_{ex,i} =\frac{J^{-}_{i} m_{i-1}}{4}+ \frac{2 J_{i} m_i}{4} + \frac{J^{+}_{i}m_{i+1}}{4}.
\end{equation}

 The spin angular momentum (normalized to $\mu_B$) at each layer is expressed as following in equation \ref{eq:equaBoltzmann}, in which $m_i$ and $S_i$ are normalized magnetizaiton (relative to its saturation magnetization) and spin quantum number at i-th layer, and $J^{-}_{i}$, $J_{i}$ and $J^{+}_{i+1}$ are exchange coupling energy per atom with the previous, current and next neighboring layer. 

\begin{multline}\label{eq:equaBoltzmann}
    m_{i} =  \frac{1}{S_i} \frac{\sum_{s = -S_{i}}^{s = S_{i}} \text{exp}(\frac{ s \, \Delta_{ex,i}}{k_B T}) }{   \sum_{s = -S_{i}}^{s = S_{i}} \text{exp}(\frac{  \, \Delta_{ex,i}}{k_B T})          }\\    
   = \frac{1}{S_i} \frac{\sum_{s = -S_{i}}^{s = S_{i}} s \, \text{exp}(\frac{ s (\frac{J^{-}_{i} m_{i-1}}{4} + \frac{2 J_{i} m_i}{4} + \frac{J^{+}_{i}m_{i+1}}{4} )  }{k_B T}) }{   \sum_{s = -S_{i}}^{s = S_{i}}  \text{exp}(\frac{s (\frac{J^{-}_{i} m_{i-1}}{4} + \frac{2 J_{i} m_i}{4} + \frac{J^{+}_{i}m_{i+1}}{4} )  }{k_B T}   )          }
\end{multline}

\begin{figure}[]
\centering
\includegraphics[width=0.9\linewidth]{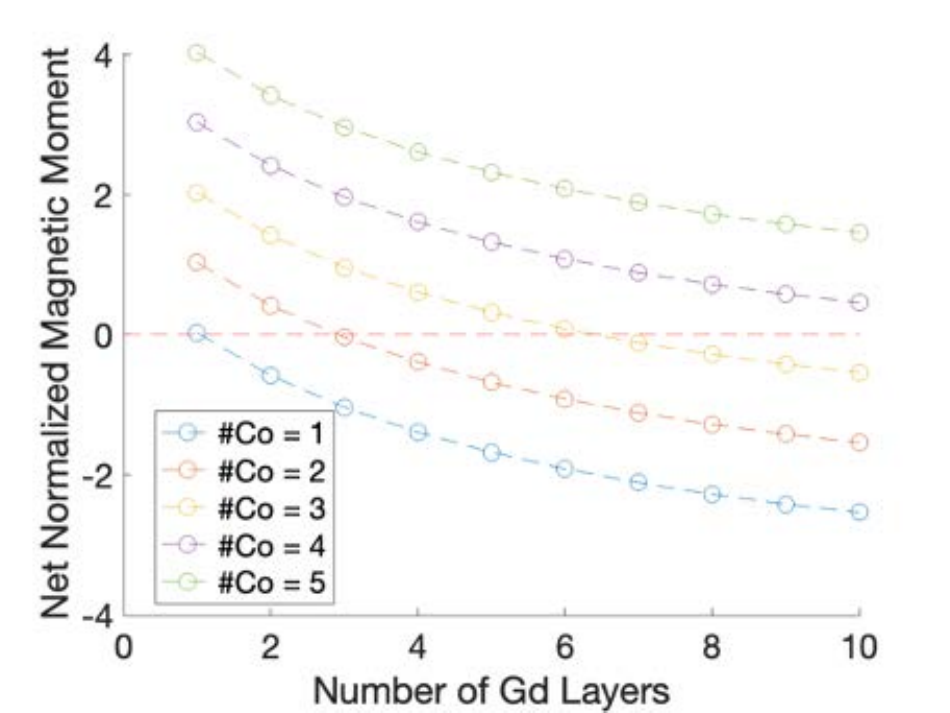}
\caption{The net normalized magnetization of Co/Gd bilayer (normalized to the saturated magnetization of single Co layer) as a function of Gd layer number. The positive sign is defined as the case in which the magnetism is Co dominated. An orange dashed line marked the compensation.)
     }
\label{fig:CurieComp}
\end{figure}

Exchange coupling at i-th layer $J_i$is defined as 

\begin{equation}\label{eq:exchangecoupling}
    J_i = \frac{3 k_B T_C^i}{S_i +1} ,
\end{equation}

whereas, the interlayer exchange between Co and Gd interface was set to be $J_{Co/Gd}/k_B = 1000 K$\cite{Coey:2010aa,Kittel:2004aa}. Here we take $T_C$ for Co and Gd as 1388 K and 292 K respectively\cite{Coey:2010aa,Kittel:2004aa}. Considering the size of the Co and Gd atom, we approximate the ratio of Gd magnetic moment per layer over that of Co to be a factor of two. We plot the normalized magnetic moment (to the magnetization of a single Co layer) of the Co/Gd bilayer as a function of underlying layer thickness in  Figure \ref{fig:CurieComp}. We found that the magnetic compensation can be obtained up to 3 Co monolayers for less than 10 layers of Gd. Our study represent the most ideal case of induced magnetism between Co and Gd.

\subsection{Sup: M3TM-based theoretical modeling of AOS with stretched pulse} \label{M3TM}

To generate the simulation results presented in Figure \ref{fig:M3TMPicoAOSSim}, illustrating the dependence of threshold energy of AOS and the thermal demagnetization on pulse duration and composition, we employed a model that simulates AOS dynamics based on M3TM. This model is also complemented by incorporating phenomenological exchange scattering in accordance with prior publications\cite{Koopmans:2010aa,Schellekens:2014aa,Beens:2019aa,Li:2022ab}.

In our simulation, temperatures are considered homogeneous over the full stack and thermal parameters of Co and Gd were taken the same. The temperature dynamics  of electronic and lattice sub-systems are modeled as:

\begin{equation}\label{eq:m3tmelectron}
    \gamma T_e \frac{dT_e}{dt} = g_{ep} (T_P - T_e) + \frac{P_0}{\sigma\sqrt{2\pi} }e^{\frac{-t^2}{2\sigma^2}}
\end{equation}

\begin{equation}
    C_p \frac{dT_p}{dt} = g_{ep}(T_e - T_p) - C_p \frac{T_p - T_{\text{amb}}}{\tau_D},
\end{equation}

in which, $T_e$ and $T_p$ are the electron and phonon (lattice) temperature, respectively, $\gamma T_e$ and $C_p$ are the electron and phonon heat capacity, $g_{ep}$ is the electron-phonon temperature coupling, $\tau_D$ is the characteristic timescale of heat dissipation ($\tau_D = $ 20 ps) and $T_{amb}$ is the ambient (room) temperature (295 K). Although full thermal relaxation can be many hundreds of ps, fast equilibration process between overheated electron reservoir with the cooler substrate can already have taken place within a few ps. Mobile electrons can pick up the energy of the fast oscillating electric field from a laser pulse. To reflect this, the laser energy dynamics is coupled as a second term in equation \ref{eq:m3tmelectron}, in which $P_0$ is the pulse energy in units of $\text{J/m}^3$ and $\sigma$ is the standard deviation of the Gaussian pulse duration. The FWHM quoted in this paper follows FWHM = $2\sqrt{2\text{ln}2} \sigma$. In our study, we assume all layer are in thermal equilibrium with respect to each other, i.e. $T_e$ and $T_p$ are uniform in all layers.

 The magnetization (in each individual layer $m_i$) in our system is modeled following a Weiss mean-free field approach (see Sup. \ref{CurieComp}) following equation \ref{eq:equaBoltzmann}. The electron temperature dynamics is directly responsible for the magnetization dynamics. A first contribution to the magnetization dynamics is Elliot-Yafet spin-flip scattering\cite{Koopmans:2010aa}, which follows 
\begin{equation}\label{eq:EY}
    \frac{dm_i}{dt}|_{\text{EY}} = R_i \frac{\Delta_{ex,i} T_p}{2k_B T_{C,i}^2} [1- m_i \text{coth}(\frac{\Delta_{ex,i}}{2k_B T_e})]
\end{equation}
in which, $\Delta_{ex,i}$ is the exchange splitting of the i-th layer and $R_i$ is the demagnetization rate of the i-th layer.  
A second contribution is the exchange scattering of the angular momentum\cite{Kittel:2004aa, Beens:2019aa}, the angular momentum exchange follows 
\begin{multline}\label{eq:ExchangeScattering}
            \frac{dm_i}{dt}|_{ex,i,i\pm1} = \frac{\eta_{i,i\pm1} \mu_B}{\mu_{at,i}} T_e^3 \\
    [-\text{SI}(\frac{-\Delta_{ex,i,i\pm1}}{k_B T_e})(\frac{1+m_i}{2})(\frac{1-m_{i\pm 1}}{2})\\
    +\text{SI}(\frac{\Delta_{ex,i,i\pm1}}{k_B T_e})(\frac{1-m_i}{2})(\frac{1+m_{i\pm 1}}{2})]
\end{multline}

where, SI is a function\cite{Beens:2019aa} integrating all the occupation level of the density of states over energy assuming a Fermi-Dirac distribution, $\mu_{at,i}$ is the atom size of the i-th layer, and $\eta_{i,i+1}$ is the exchange scattering rate between i-th layer and i+1-th layer. 

The magnetization dynamics of the i-th layer involving the mentioned contribution can be summarized as follows
\begin{equation}
    \frac{dm_i}{dt} = \frac{dm_i}{dt}|_{\text{EY}} + \frac{dm_i}{dt}|_{ex,i,i+1} +\frac{dm_i}{dt}|_{ex,i,i-1}
\end{equation}

\begin{figure*}[]
\centering
\includegraphics[scale=0.2]{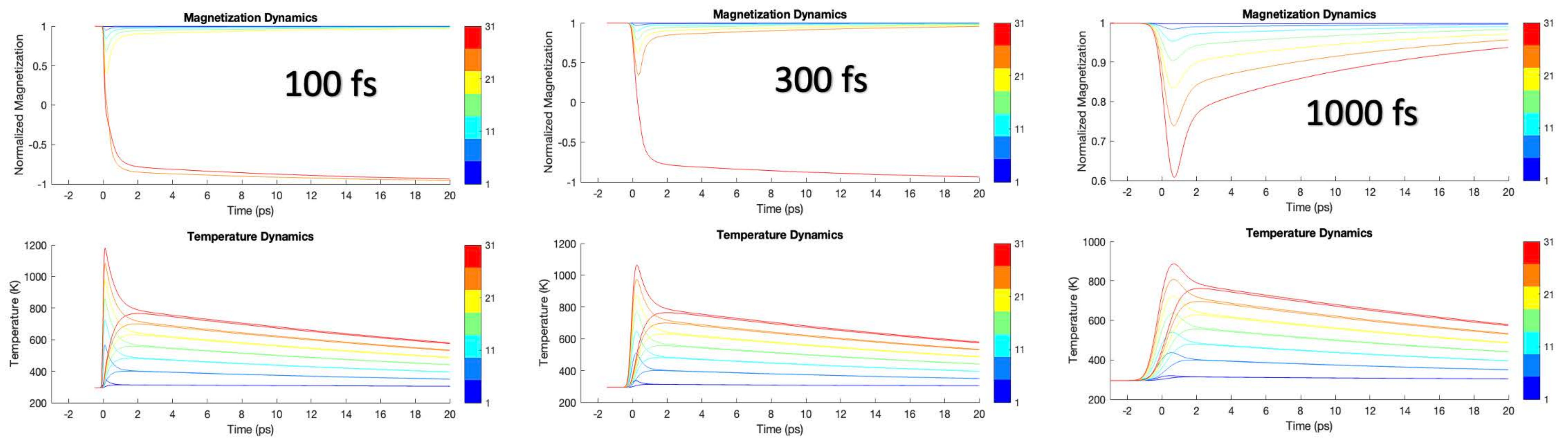}
\caption{The simulated AOS dynamics (up) and two-temperature evolution (electron and phonon) (down) of Co(2L)/Gd(4L) structure excited by laser pulses of standard deviation $\sigma$ = 100 fs, 300 fs and 1000 fs respectively (note that FWHM = $2\sqrt{2\text{ln}(2)}\sigma$) as a function of pulse energy. The electron temeprature and phonon temperature is represented in the same color, but can be differentiated by the underlying behavior as electron temperature can temporarily rise up a lot higher than phonon temperature. The pulse energy (in in unit of $10^8$ J/m$^3$) is presented in the color bar. 
     }
\label{fig:M3TMDynamics}
\end{figure*}

Previous work following the above-mentioned identical approach, predicts an AOS dynamic with a dramatic "plateau", i.e. the magnetization stuck at 0 for significant long time, such dynamics is however not experimentally observed(see ref. \cite{Peeters:2022uo} and results in  Figure \ref{fig:Fig1}g-h). 
We understand this behavior as the strength in exchange scattered angular momentum is not sufficient while the electron reservoir is sufficiently heated above the Curie temperature. Moreover, keeping the original parameters\cite{Schellekens:2013aa,Beens:2019aa, Koopmans:2010aa}, the resulting threshold energy for AOS is much larger than experimentally quantified value and the energy window (between AOS energy and Curie energy) is very little compared with our experiment of Co/Gd multilayer (see  Figure \ref{fig:ExperimentPicoAOS}). Given that the exact report on the value of exchange scattering efficiency is lacking and the mechanism is highly phenomenological, we increased the exchange scattering strength between Co and Gd ($\eta_{\text{Co,Gd}}$) by a factor 10 and reduced the spin-flip probability of Gd ($R_{\text{Gd}}$) by a factor of two (while keeping the rest of the parameters as identical as in ref. \cite{Koopmans:2010aa,Schellekens:2014aa,Beens:2019aa}), in this way, we can significantly reduce the "plateau"\cite{Beens:2019aa} (see Figure \ref{fig:M3TMDynamics}) to match the dynamic behaviour similar to our experimental results presented in Figure \ref{fig:Fig1}g-h qualitatively, and also create a much larger energy window\cite{Beens:2019aa} as similar to our experiment.

The parameters mentioned in above discussion are presented in the table below

\begin{table}[h]
\begin{center}
\begin{minipage}{174pt}
\label{Table:ParametersTable}%
\begin{tabular}{@{}lll@{}}
\toprule
Parameter & Value  &  Unit\\
\midrule
$C_\mathrm{p}$    & 4$\times$10$^{6}$   & Jm$^{-3}K^{-1}$   \\
$g_\mathrm{ep}$    & 4.05$\times$10$^6$   & Jm$^{-3}$K$^{-1}$ps$^{-1}$    \\
$\gamma $     & 2.0$\times$10$^3$   & Jm$^{-3}$K$^{-2}$  \\
$\tau_\mathrm{D}$    & 20   & ps   \\
$R_\mathrm{Co}$    & 9.550$\times$10$^5$   & Am$^{-1}$ps$^{-1}$   \\
$R_\mathrm{Gd}$     & 0.196$\times$10$^5$   & Am$^{-1}$ps$^{-1}$   \\
$T_\mathrm{C,Co}$    & 1388   & K   \\
$T_\mathrm{C,Gd}$    & 292   & K    \\
$\mu_\mathrm{at, Co}$    & 1.72   & $\mu_B$   \\
$\mu_\mathrm{at, Gd}$    & 7.55   & $\mu_B$   \\
$S_\mathrm{Co}$    & 1/2   &    \\
$S_\mathrm{Gd}$    & 7/2   &    \\
$\eta_\mathrm{Co-Co}$    & 4.615   & ns$^{-1}$   \\
$\eta_\mathrm{Gd-Gd}$    & 0.012   & ns$^{-1}$   \\
$\eta_\mathrm{Co-Gd}$    & 0.554   & ns $^{-1}$ \\
\botrule
\end{tabular}
\end{minipage}
\end{center}
\caption{Summary of the parameters in equations shown in Sup. \ref{CurieComp} and \ref{M3TM}.}
\end{table}

We now present some characteristic results of the dynamics of AOS using pulses with different pulse durations (see  Figure \ref{fig:M3TMDynamics}). In this case, we used pulses identical in pulse energy ($P_0$) but differ in pulse duration ($2\sqrt{2ln2} \sigma$). We see, in  Figure \ref{fig:M3TMDynamics}, as $\sigma$ increases from 0.1 ps to 1 ps, the dynamics gradually transits from AOS switching dynamics towards demagnetization dynamics. In the meantime, the transient difference between electron and phonon temperature reduces. We interpret this behavior as the strength of exchange scattering and spin-flip scattering are monotonically related to the non-adiabatic temperature difference and the peak electron temperature. As the pulse is prolonged, especially closer to the characteristic timescale of electron-phonon coupling\cite{Koopmans:2010aa}, the phonon temperature can more easily follow the electron temperature, resulting a low temperature difference and low peak temperature. Our results here also qualitatively matches that of atomistic-LLB\cite{Jakobs:2021aa, Raposo:2022aa,Wei:2021ui}. 

\subsection{Sup: Determination of threshold energy for AOS and thermal demagnetization from M3TM-based theoretical modeling} \label{M3TMphase1}

\begin{figure*}
\centering
\includegraphics[width=1.0\linewidth]{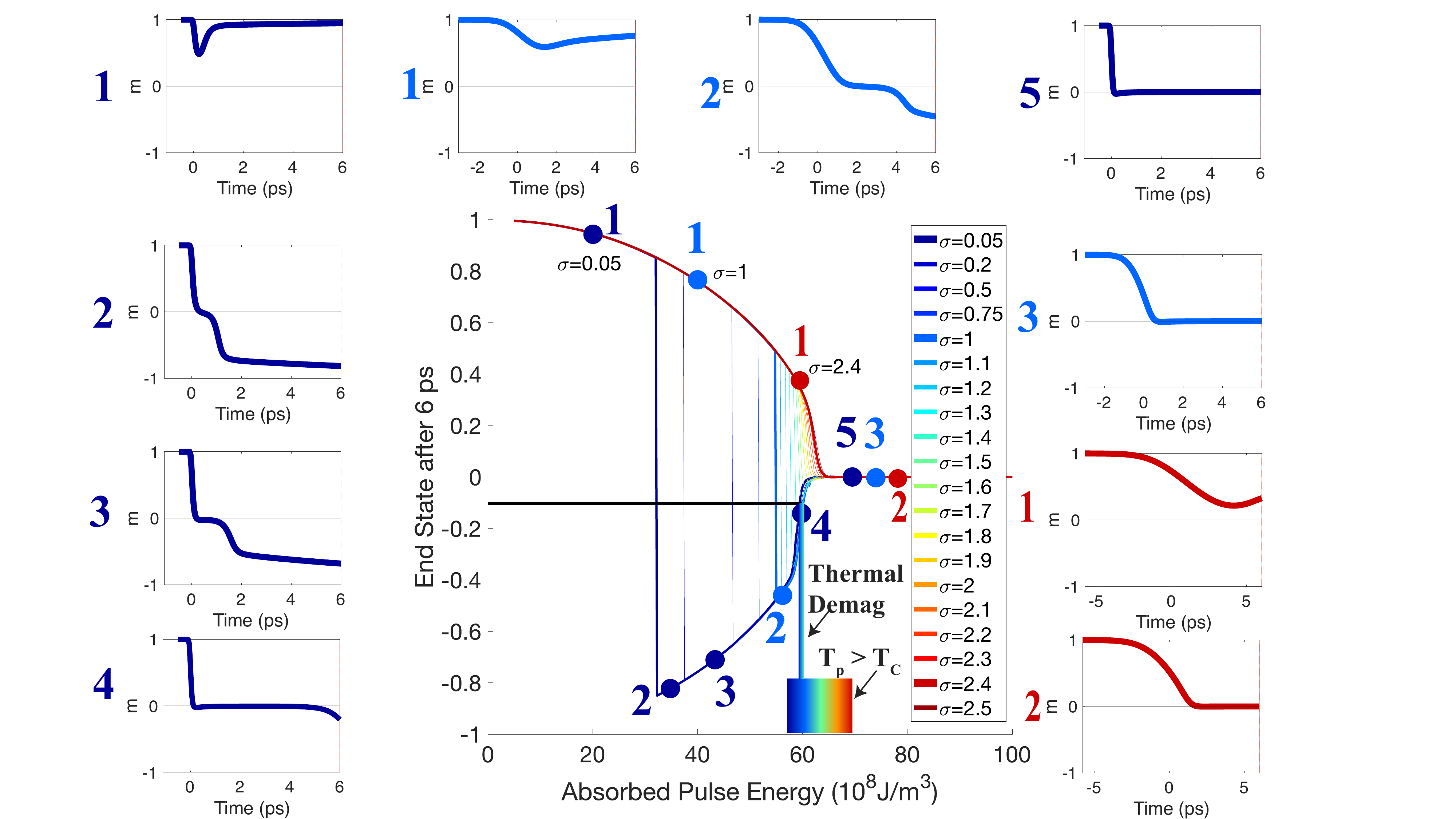}
\caption{   (Center) A plot of the normalized magnetization of Co component (averaged over all Co layer) from Co(4L)/Gd(4L) obtained at the 6 ps of its dynamics after the pulse illumination (peak reached at time 0 ps) as a function of standard deviation $\sigma$ of the Gaussian pulse and pulse energy (in unit of $10^8$J/m$^3$). -10$\%$ of the end state after 6 ps is also marked using a horizontal black line, which is used to differentiate cases between switching and thermal demagnetization. The threshold pulse energy for thermal demagnetization, as the intersection between the black line and the plots with different pulse duration, is also marked up as a vertical colorbar,  which also indicates the threshold energy for different pulse duration (expressed in $\sigma$). Note that the pulse duration (FWHM) of the gaussian pulse, as quoted in our paper, is expressed as FWHM = $2\sqrt{2ln2} \sigma$. Some examples of magnetization dynamics of pulses with $\sigma$ = 0.05 ps (dark blue), $\sigma$ = 1.0 ps (light blue) and $\sigma$ = 2.4 ps (red) are presented along side the plot. The pulse energy and pulse duration of these selected values for pulse duration and energy are also marked up in the center plot. The threshold pulse energy for different pulse duration (expressed in $\sigma$), at which the peak phonon temperature during its dynamics exceeds Curie temperature, is marked up as a vertical color bar. }
\label{fig:M3TMPhase1}
\end{figure*}

Now we proceed to extract the threshold energy for AOS and the maximum pulse duration from the M3TM-based simulations mentioned earlier. To illustrate such a process, we focus on the composition of Co(4L)/Gd(4L) as a characteristic example. 
Figure \ref{fig:M3TMPhase1} presents a plot of the magnetization at 6 ps (note that the pulse peak occurs at 0 ps) for different pulse energies and durations.

When using a pulse with $\sigma$ = 0.05 ps and low energy, we observe a transient quenching of magnetization followed by a remagnetization process. This demagnetization process is also evident with a longer pulse ($\sigma$ = 1 ps in  Figure \ref{fig:M3TMPhase1}), but at a higher pulse energy. As the pulse energy surpasses a certain threshold, AOS in magnetization occurs. We define this threshold as the threshold energy for AOS.

With further increases in energy, AOS continues to occur, but with an extended plateau (see  Figure \ref{fig:M3TMPhase1}) and a less pronounced end state on the opposite side. Our model attributes this behavior to the fact that overheating reduces magnetization, increases spin-flip probability, and decreases exchange scattering rate, as described by equations \ref{eq:EY} and \ref{eq:ExchangeScattering}. 

As the pulse energy is raised even higher, the final magnetization state approaches zero or very close to it (see  Figure \ref{fig:M3TMPhase1}).  In these cases, magnetization remains zero after losing its angular momentum to the lattice without crossing zero, or crossing 0 after staying at 0 for a prolonged time  (see  Figure \ref{fig:M3TMPhase1}). It is also worth noting that, in these cases, the peak phonon temperature is already very close to Curie temperature (see Figure \ref{fig:M3TMPhase1}). Specifically, with short pulses, the peak phonon temperature is already higher than the Curie temperature, whereas with long pulses, the peak phonon temperature can be slightly lower. Additionally, it's evident that as pulse duration increases, zero crossing becomes less likely until it eventually disappears. This aligns with the established behaviors of AOS processes at longer pulse durations, both experimentally and theoretically, as demonstrated in atomistic-LLG \cite{Jakobs:2021aa, Raposo:2022aa}. We attribute this behavior to the well-documented thermal demagnetization processes commonly associated with multi-domain states, as observed in numerous experimental studies\cite{Stanciu:2007aa, Ostler:2012aa, Lalieu:2017aa, Wei:2021ui, Zhang:2022aa}.

In our studies, we chose to define the threshold energy for thermal demagnetization from our model based on the criteria that, the end state at 6 ps obtained from our simulation should be at least lower (more negative) than -10$\%$ to be able to qualify as AOS. This choice is motivated by the fact that, thermal fluctuation field (which is not implemented in our model) can be expected to easily interrupt a long dwelling at zero. As such a fluctuation introduces randomistic direction of small angular momentum, which is enough to break the degeneracy at zero developing to a undeterministic state in reality.
We, thus, consider the switching processes with dwelling time longer than 6 ps or a slight switching more positive than -10$\%$ as also thermal demagnetization.

\subsection{Sup: AOS in [Co/Gd]$_n$ by a 50 fs laser pulse}\label{CoGdBilayer}

\begin{figure*}
\centering
\includegraphics[width= \linewidth]{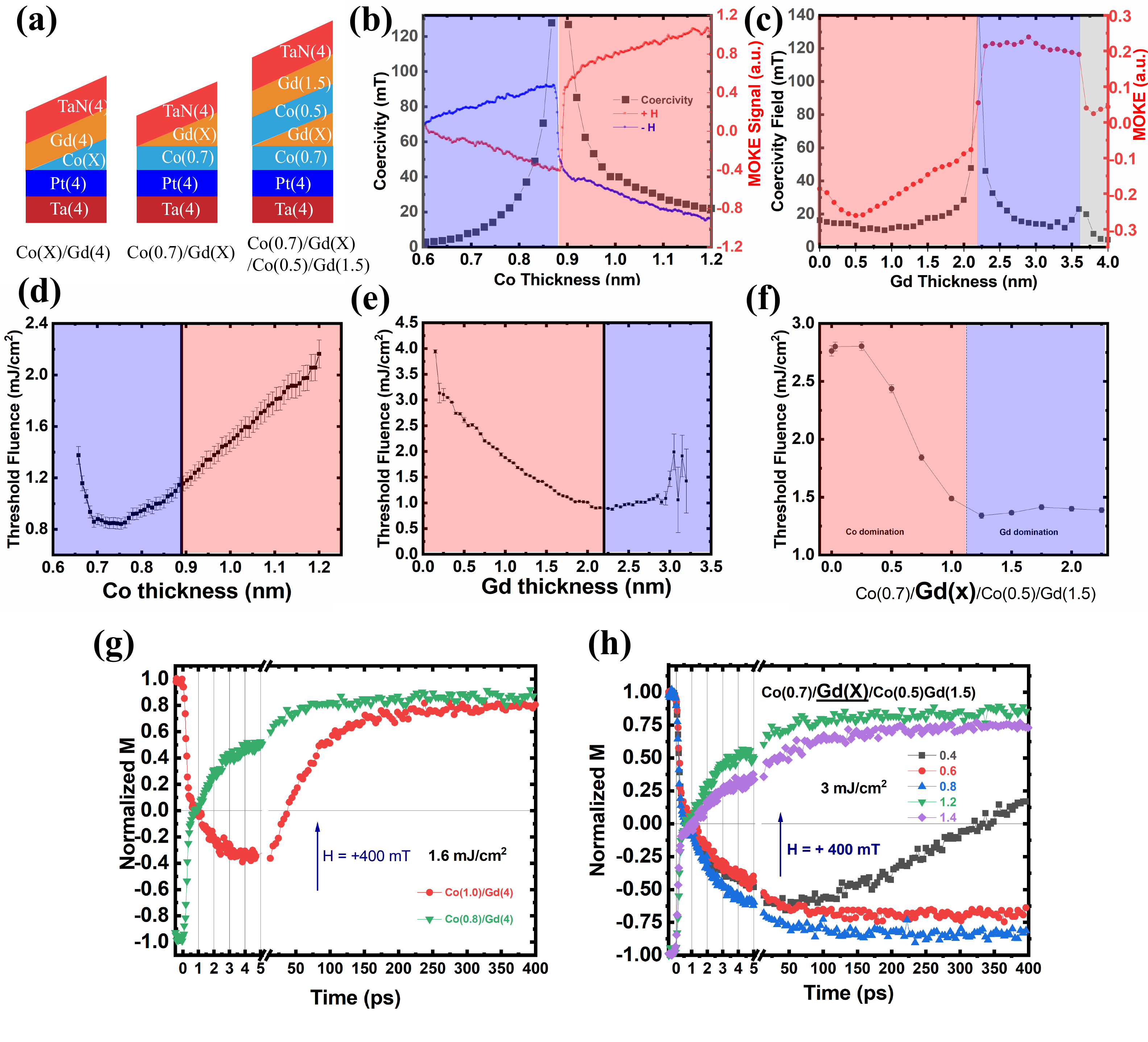}
\caption{(a) The schematic illustration of the wedge sample stacks including Co/Gd bilayer Co(X)/Gd(4), Co(0.7)/Gd(4) and [Co/Gd]$_2$. The substrate Si/SiO$_{100}$ of the all the sample is not present in the illustration. 
    (b) - (e) include the wedge scan study of magneto-statics characterized by MOKE and threshold fluence of AOS using a 50 fs laser pulse. Here, we note that, due to the large spot size of our probe laser, inaccuracy of 5$\%$ as the absolute position error bar should be expected (not present in the figure).  
    (b) MOKE signal of Bi:Co, where as the signal of the sample saturated at +H (red) and -H (blue), and Coercivity (black) as a function of Co thickness.  
    (c) MOKE signal of Bi:Gd, where as the signal of the sample saturated at +H (red) and Coercivity (black) as a function of Gd thickness.    
    (d)-(e)-(f)  Threshold fluence (by 50 fs laser pulse) of Bi:Co, Bi:Gd and Quad:Gd as a funciton of wedge layer thickness. The magnetic compensation is marked in the figure as a vertical line. 
    (g)-(h) The switching dynamics of various composition (marked in the legend) of Bi:Co and Quad:Gd. The incident fluence is marked in the figure. Magnetization characterized using TRMOKE is normalized, the M>0 corresponds to the direction of the Co magnetic moment parrallel to the externally applied magnetic field.  
     }
\label{fig:Fig1}
\end{figure*}

In our study, additional samples as represented in Figure \ref{fig:Fig1}a were created to give further information on the generic behaviour of magnetostatics of Co/Gd bilayers, and also AOS statics and dynamics of [Co/Gd]$_n$ using a 50 fs laser pulse. The samples were created following the method presented in our main paper.

First, we present the magnetic statics of Co/Gd bilayer, whose ps-AOS behavior are presented in Figure \ref{fig:ExperimentPicoAOS} in the main text.
We characterized the magnetic statics of the presented samples using polar-MOKE ellipticity at wavelength of 658 nm. In particular, we measured the coercivity and the MOKE contrast as a function position (composition) by sweeping magnetic field along the out of plane direction at the same rate. The positive sign of MOKE signal corresponds to the direction of the Co magnetic moment aligned with the direction of the external magnetic field. 
We present our results on bilayer wedges of Co(0.6-1.2)/Gd(4) and Co(0.7)/Gd(0-4).
In Figure \ref{fig:Fig1}b, we plot the coercivity and MOKE signal of Co(0.6-1.2)/Gd(4). 
At the Co thickness of 0.88 nm, the bilayer is at magnetization compensation, which is evidenced by a divergent coercivity and flipping sign of MOKE signal. Below this compensation thickness, the system is in Gd dominated regime.
Similarly, for Co(0.7)/Gd(0-4) (see  Figure \ref{fig:Fig1}c), the Gd dominated regime is reached at a Gd thickness above 2.2 nm. The existence of Gd dominated regime at room temperature, to our knowledge have not been reported in Co/Gd bilayer system\cite{Pham:2016aa,BlaesingECT2018,Lalieu:2017aa,Lalieu:2019aa,Tan:2019wf}. It is worth noting that the Co(0.7)/Gd(2.1) presented in our main paper (see  Figure \ref{fig:ExperimentPicoAOS}) is Gd dominated. The underlying difference, i.e. the difference in compensation thickness, we believe, lies in the sample to sample thickness variation, uncertainties in the thickness calibrations of both Co and Gd.

Our theoretical calculation, based on Weiss mean-field model, shows that compensation could not be obtained with more than 2 monolayers of Co from a pristine Co/Gd interface (see Sup. \ref{fig:CurieComp}). Thus, we also consider the interfacial intermixing between Co and Gd, which can potentially induce more magnetization of Gd, as another contributing mechanism leading to the room temperature compensation.

Next, we present our AOS threshold fluence using 50 fs linearly polarized laser pulses as a function of composition. 
We first present the AOS threshold fluence of Co(0.6-1.2)/Gd(4), Co(0.7)/Gd(0-4) and Co(0.7)/Gd(0-2.2)/Co(0.5)/Gd(1.5), which is shown in  Figure \ref{fig:Fig1}d-f. As for Co(0.6-1.2)/Gd(4) (see \ref{fig:Fig1}d), it can be observed that the threshold fluence decreases with Co thickness, which is consistent with our theoretical calculation (see  Figure \ref{fig:M3TMPicoAOSSim}) and our experimental result (see  Figure \ref{fig:ExperimentPicoAOS}) due to the reduction of Curie temperature\cite{Beens:2019aa,Jakobs:2021aa,Lalieu:2017aa,Li:2021wr}. Interestingly, crossing compensation to the Gd domination (at thinner Co), it can be seen that the trend of decrease is flattened and eventually tilt up. As for Co(0.7)/Gd(0-4)(see \ref{fig:Fig1}e), as Gd thickness increases from the lowest thickness of Gd (0.2 nm) enabling the existence of AOS, the AOS threshold fluence decreases by a factor of four at the Co dominated regime and levels up in in the Gd dominated regime. The behavior in Gd dominated regime is similar to the experimental results of that for [Co/Gd]$_2$ for Gd thickness beyond compensation (see Figure \ref{fig:ExperimentPicoAOS}d and a zoomed in view in  Figure \ref{fig:Fig1}f). 
As for qualdyaer, consistently with previous report\cite{Li:2022ac}, the threshold fluence decreases as Gd thickness increases, reaching a minimum at magnetization compensation and remains almost unchanged at Gd domination at even thicker Gd. A common feature can be briefly summarized here, using a 50 fs laser pulse, higher Gd content decreases AOS threshold fluence at Co domination, while at Gd domination, the AOS energy efficiency is suppressed.

Having demonstrated the static AOS, we now switch gear to the AOS dynamics of [Co/Gd]$_2$. The dynamics measurement follows identical protocol as in previous works\cite{Xu:2017aa,Zhang:2022aa}. Here we denote the positive sign of MOKE signal as the direction of Co magnetic moment is parallel with the external magnetic field applied during the measurement. 
As for bilayer Co/Gd (see  Figure \ref{fig:Fig1}g), it can be seen that the magnetization of Co crosses 0 at less than one ps as a result of a strong exchange scattering. While at longer timescale (50-400 ps), the Co dominated sample was able to reset its magnetization. On the other hand, for the Gd dominated sample, the switching  continues and was able to reset at a much longer timescale. We attribute this effect to the presence of an external magnetic field. As cobalt switches its sign, its magnetization becomes aligned with the external magnetic field. The fact that the Co magnetization in both sample follows external magnetic field, indicates both systems are in Co dominated regime at the timescale of 50-400 ps. 
Similar behaviors are observed in Co(0.7)/Gd(X)/Co(0.5)/Gd(1.5) (see Figure \ref{fig:Fig1}h). Nevertheless, as the composition is getting closer to the compensition (Gd thickness of 1.1 nm), the magnetization dynamics at longer timescale becomes less susceptible to the magnetic field compared with bilayer case. This means the system is closer to magnetization compensation. 

\end{document}